\documentclass[aps,prb,twocolumn,superscriptaddress,floatfix]{revtex4}

\usepackage{amsmath,amssymb,bm}
\usepackage{epsfig}
\usepackage{graphics}

\newcommand{\bk}{{\bf k }}

\newcommand{\br}{{\bf r }}

\newcommand{\bv}{{\bf v }}

\newcommand{\bG}{{\bf G}}

\newcommand{\bK}{{\bf K}}

\begin{document}

\title{Anisotropic behaviors of massless Dirac fermions in graphene under periodic potential}
\author{Cheol-Hwan Park}
\affiliation
{Department of Physics, 
University of California at Berkeley, 
Berkeley, California 94720, USA}
\affiliation
{Materials Sciences Division, 
Lawrence Berkeley National Laboratory, Berkeley, 
California 94720, USA}
\author{Li Yang}
\affiliation
{Department of Physics, 
University of California at Berkeley, 
Berkeley, California 94720, USA}
\affiliation
{Materials Sciences Division, 
Lawrence Berkeley National Laboratory, Berkeley, 
California 94720, USA}
\author{Young-Woo Son}
\affiliation
{Department of Physics, 
Konkuk University,
Seoul, Seoul 143-701, Korea}
\author{Marvin L. Cohen} 
\affiliation
{Department of Physics, 
University of California at Berkeley, 
Berkeley, California 94720, USA}
\affiliation
{Materials Sciences Division, 
Lawrence Berkeley National Laboratory, Berkeley, 
California 94720, USA}
\author{Steven G. Louie}
\affiliation
{Department of Physics, 
University of California at Berkeley, 
Berkeley, California 94720, USA}
\affiliation
{Materials Sciences Division, 
Lawrence Berkeley National Laboratory, Berkeley, 
California 94720, USA}
\date{{\it submitted}, October 2, 2007}
\maketitle
{\bf Charge carriers of graphene
show neutrino-like linear energy dispersions
as well as chiral behavior near
the Dirac point~\cite{wallace:1947PR_BandGraphite,
novoselov:2005PNAS_2D,novoselov:2005Nat_Graphene_QHE,
zhang:2005Nat_Graphene_QHE,
berger:2006Graphene_epitaxial,
geim:2007NatMat_Graphene_Review}.
Here we report
highly unusual
and unexpected behaviors of these carriers
in applied external periodic
potentials, i.e., in graphene superlattices.
The group velocity
renormalizes
highly anisotropically
even to a degree that
it is not changed at all
for states with wavevector
in one direction but is reduced to zero in another,
implying the possibility that
one can make nanoscale electronic circuits out of
graphene not by cutting
it~\cite{son:2006GNR_LDA,son:2006GNR_Halfmetal,
han:2007PRL_GNR_bandgap,chen:2007condmat_GNR_bandgap}
but by drawing on it in a non-destructive way.
Also, the type of charge carrier species
(e.g. electron, hole or open orbit)
and their density of states
vary drastically with the Fermi energy,
enabling one to tune the Fermi surface-dominant
properties significantly with gate voltage.
These results address the fundamental
question of how chiral massless Dirac fermions
propagate in periodic potentials and point to a new
possible path for nanoscale electronics.}

Since the pioneering work by
Esaki and Tsu~\cite{esaki:1970IBM_Superlattice},
superlattices have been studied extensively
and have had a huge impact on semiconductor
physics~\cite{tsu:2005SL_to_NEs,cottam:1989Intro_SL}.
Superlattices demonstrate interesting phenomena
such as negative differential conductivity,
Bloch oscillations, gap openings at the
mini Brillouin zone boundary formed by the additional
periodic potential,
etc~\cite{tsu:2005SL_to_NEs,cottam:1989Intro_SL}.
Conventional semiconducting superlattices are mainly produced
by molecular-beam epitaxy and metallo-organic
chemical vapour-phase deposition while metallic superlattices
are made by sputtering
procedures~\cite{tsu:2005SL_to_NEs,cottam:1989Intro_SL}.
We expect that, by modulating the potential seen by the electrons,
graphene superlattices may be fabricated by adsorbing adatoms on
graphene surface through similar techniques,
by positioning and aligning impurities with
scanning tunneling microscopy~\cite{eigler:1990STM_IBM,
crommie:1993STM_corral,hiura:2004ASS_Graphite_STM},
or by applying a local top-gate voltage to
graphene~\cite{williams:2007Sci_Graphene_topgate,
huard:2007PRL_Graphene_Transport,ozyilmaz:2007PRL_Graphene_Transport}.
Epitaxial growth of graphene~\cite{berger:2006Graphene_epitaxial}
on top of pre-patterned substrate is also a possible route
to graphene superlattice. Recently, periodic pattern in the
scanning tunneling microscope image
has been demonstrated on a graphene monolayer on top of
a metallic substrate~\cite{marchini:2007PRB_Graphene_Ru,vazquez:2008PRL_Graphene_SL,
pan:2007condmat_Graphene_SL} as well.

  \begin{figure}
  \includegraphics[width=0.95\columnwidth]
  {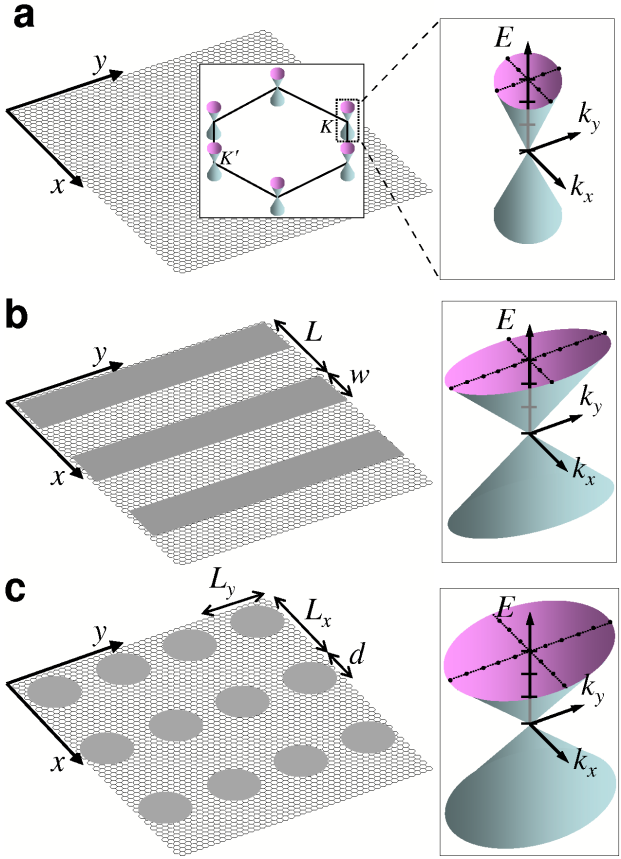}
  \caption{{\bf Graphene superlattices and anisotropic Dirac cones.}
  {\bf a}, Schematic diagram of graphene. Inset: the Brillouin zone
  of graphene and Dirac cones centered at Dirac points among which two
  (K and K') are nonequivalent (left) and the linear and
  isotropic energy dispersion near one of the Dirac points
  of charge carriers in graphene (right).
  {\bf b}, One-dimensional (1D) graphene superlattice
  formed by Kronig-Penney type
  of potential
  periodic along $\hat x$ direction with spatial period $L$ and barrier width
  $w$.
  The potential is $U_{\rm 1D}$ in the grey regions and zero outside.
  Inset: energy dispersion of charge carriers in 1D graphene superlattice.
  The energy dispersion along any line in two-dimensional (2D) wavevector
  space from the Dirac point is linear but with different group velocity.
  For a particle moving parallel to the
  periodic direction, the group velocity ($v_{\parallel}$) is
  not renormalized at all whereas that for a particle moving
  perpendicular to the periodic direction ($v_{\perp}$) it is reduced most.
  {\bf c}, 2D graphene superlattice with muffin-tin type of
  potential periodic along both $\hat x$ and $\hat y$ directions
  with spatial periods $L_x$ and $L_y$, respectively. The potential is
  $U_{\rm 2D}$ inside the grey disks with diameter $d$ and zero outside.
  Inset: energy dispersions of charge carriers in 2D graphene superlattice.}
  \label{Fig1}
  \end{figure}

The low energy charge carriers in pristine graphene
are described by a massless Dirac
equation and have a linear energy dispersion which is
isotropic near the Dirac points K and K'
in the Brillouin zone~\cite{wallace:1947PR_BandGraphite,ando:1998JPSJ_NT_Backscattering,
ando:1998JPSJ_NT_BerryPhase,novoselov:2005Nat_Graphene_QHE,zhang:2005Nat_Graphene_QHE,
divincenzo:1984PRB_GIC_EffMass} (Fig.~1a).
It is shown experimentally that the carriers have a group velocity of
$v_0\approx10^6$~m/s which plays the role of an effective speed
of light in (2+1)
dimensional quantum electrodynamics~\cite{novoselov:2005Nat_Graphene_QHE,zhang:2005Nat_Graphene_QHE}.
Within the effective-Hamiltonian approximation,
$$
H=\hbar v_0\left(
\begin{array}{cc}
0 & -ik_x-k_y\\
ik_x-k_y & 0
\end{array}
\right)\ ,
$$
where $k_x$ and $k_y$ are the $x$ and $y$ components of
the wavevector $\bk$ of the Bloch state
defined with respect to the Dirac point,
the wavefunction of the quasiparticles in graphene
has two components corresponding to the amplitude on the
two different trigonal sublattices of graphene and can be represented by
a two component spinor~\cite{divincenzo:1984PRB_GIC_EffMass,
ando:1998JPSJ_NT_Backscattering,ando:1998JPSJ_NT_BerryPhase}.
This spinor structure of the wavefunction is called a pseudospin
(because it is not related to a real spin)
or chirality~\cite{divincenzo:1984PRB_GIC_EffMass,
ando:1998JPSJ_NT_Backscattering,ando:1998JPSJ_NT_BerryPhase,
mceuen:1999PRL_NT_Backscattering,katsnelson:2006NatPhys_Graphene_Klein},
which is of central importance to the novel physical properties of
graphene superlattices discussed below.

Let us now consider the situation that an additional periodic
potential is applied to graphene.
If the spatial period of the superlattice potential is
much larger than the nearest neighbor carbon-carbon distance in graphene
($\sim1.42$~\AA), the scattering of a state close to one Dirac point
to another one does not occur~\cite{ando:1998JPSJ_NT_Backscattering,
ando:1998JPSJ_NT_BerryPhase,mceuen:1999PRL_NT_Backscattering}.
Therefore, even though there are two nonequivalent Dirac cones
for the energy dispersion surface of graphene,
focusing on a single cone is sufficient.
This condition also implies that, in the graphene superlattices discussed here,
there is no gap opening at the Dirac point~\cite{ando:1998JPSJ_NT_Backscattering,
ando:1998JPSJ_NT_BerryPhase,mceuen:1999PRL_NT_Backscattering}.

To investigate the physics of
charge carriers in graphene superlattices,
we have calculated the energy dispersions, the group velocities,
and energy gap openings
at the minizone boundaries (MB)
within the effective-Hamiltonian formalism~\cite{wallace:1947PR_BandGraphite}.
Effects of the external periodic potential are incorporated into our calculation
through the scattering matrix elements between pseudospin states, or chiral
eigenstates, of the electrons in graphene:
$$
\left<s,\bk\left| U(\br)\right|s',\bk'\right>
=\sum_{\bG}\frac{1}{2}\left(1+ss'e^{-i\theta_{\bk,\bk-\bG}}\right)U(\bG)\ 
\delta_{\bk',\bk-\bG}\ .
$$
Here, $\bG$'s are reciprocal lattice vectors of the superlattice
and $\theta_{\bk,\bk-\bG}$ is the angle between $\bk$ and $\bk-\bG$.
$U(\br)$ and $U(\bG)$ are external potential in real and wavevector
space, respectively. $s$ and $s'$ are either $+1$ or $-1$ depending on
whether the energy of the state is above or below
the energy at the Dirac point, respectively.
We have also carried out a tight-binding formulation and obtained
identical results as those discussed below.

  \begin{figure}
  \includegraphics[width=1.0\columnwidth]
  {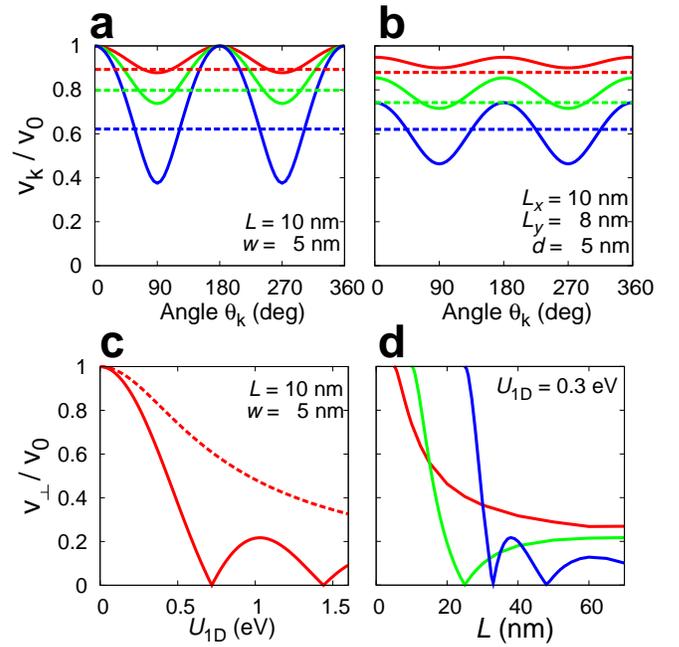}
  \caption{{\bf Anisotropic velocity renormalization in graphene superlattices.}
  {\bf a}, The component of the group velocity
  parallel to the $\bk$ vector
  [$v_{\hat k}\equiv\bv(\bk)\cdot\hat k$ with $\bk$ measured from the Dirac point]
  of charge carriers in a 1D graphene
  superlattice in units of the Fermi velocity in graphene ($v_0$)
  versus the angle ($\theta_\bk$) of the $\bk$-vector
  from the periodic potential direction $\hat x$
  (solid lines) and that in a superlattice made from a fictitious
  system of nonchiral fermions with properties otherwise identical
  to those in graphene (dashed lines).
  Red, green and blue lines correspond to $U_{\rm 1D}$ being
  0.2~eV, 0.3~eV and 0.5~eV, respectively.
  {\bf b}, Similar quantities as in {\bf a} for a rectangular 2D graphene
  superlattice. Red, green and blue lines correspond to $U_{\rm 2D}$ being
  0.3~eV, 0.5~eV and 0.7~eV, respectively.
  {\bf c},
  The group velocity of charge carriers in a 1D graphene superlattice (solid line)
  with $\bk$ perpendicular to
  the periodic direction, $v_{\perp}$, in units of $v_0$
  versus $U_{\rm 1D}$ (solid line) and that in a superlattice made
  from a fictitious
  system of nonchiral fermions with properties otherwise identical
  to those in graphene (dashed line).
  {\bf d},
  $v_{\perp}$ versus the potential spatial period ($L$)
  of charge carriers in a 1D graphene superlattice.
  Red, green and blue lines correspond to
  a fixed potential barrier height but with width ($w$)
  being 5~nm, 10~nm and 25~nm, respectively.}
  \label{Fig2}
  \end{figure}

First, for a one-dimensional (1D) graphene superlattice (Fig.~1b),
we find that the group velocity for states with wavevector $\bk$
($\bk$ is the wavevector of the Bloch state
defined with respect to the Dirac point),
is anisotropically renormalized, i.e., it is a strong function
of the direction of $\bk$.
For pristine graphene, the group velocity of states
near the Dirac point is parallel to $\bk$ and of constant
magnitude ($v_0$).
For example, in a 1D superlattice of Kronig-Penny type
of periodic potential with potential barrier height
($U_{\rm 1D}$) of 0.5~eV and
spatial period ($L$) and barrier width ($w$) of 10~nm
and 5~nm, respectively, the group velocity of the charge carriers
when $\bk$ is
along certain direction is renormalized to be slower than 40~\%
of its original value $v_0$ but is the same as $v_0$
along some other direction. [Fig.~2a:
the plotted quantity $v_{\hat k}\equiv\bv(\bk)\cdot\hat k$
is the component of the group velocity
parallel to the wavevector $\bk$ in units of $v_0$.
We note that this quantity which depends only on the
direction of $\bk$ (Supplementary Discussion 2)
is exactly equal to the absolute value of
the group velocity $v_g$ when
$\bk$ is at 0, 90, 180 or 270 degrees from
the periodic direction of the applied potential
and, when the applied potential is weak,
is only slightly different from $v_g$
at other angles (Supplementary Discussion 3).]
Thus, the group velocity of charge carriers can be tailored
highly anisotropically in graphene superlattices.
More interestingly, the group velocity
when $\bk$ is along the direction perpendicular to the periodic
direction of the potential ($v_\perp$)
is reduced the most, whereas when $\bk$ is in the parallel direction,
it is not renormalized at all (Fig.~1b).
This result is counter-intuitive since the velocity is
strongly reduced when the charge carrier is moving
parallel to the hurdles,
but is not modulated when it is crossing them.

To understand the physics behind this phenomenon,
we have performed the same calculation for a fictitious system
with carriers that have no chirality but otherwise
identical to those in graphene including the linear energy dispersion.
The group velocity in this system is reduced
isotropically and the renormalized group velocity is
close to $v_{\perp}$, i.e., the maximally renormalized one,
in 1D graphene superlattices (Fig.~2a).
Thus, it is clear that the absence of velocity renormalization in the direction
parallel to the periodicity of the external potential originates from
the chiral nature of the electronic states of graphene.
This behavior can be demonstrated more clearly by second order
perturbation theory in the case of the 1D periodic potential
with weak amplitudes (Supplementary Discussion~2).
We note that the chirality discussed here is also important in
tunneling phenomenon in graphene
through a single barrier~\cite{katsnelson:2006NatPhys_Graphene_Klein}
or a finite number of barriers~\cite{bai:2007PRB_Graphene_SL}.

In the case of two-dimensional (2D) graphene superlattices,
the group velocity is renormalized for $\bk$ states
along every direction, but anisotropically (Fig.~1c).
As the amplitude of the potential increases
the overall group velocity is reduced
and the ratio of the maximum group velocity to the minimum one
is enhanced (Fig.~2b).
Here, again, the anisotropy disappears if the chiral nature of the states
in graphene is arbitrarily removed.
As was demonstrated for the 1D superlattice,
the sinusoidal dependence on the angle of propagation
as well as the overall shift in the case of 2D graphene superlattice
of the component of the
renormalized group velocity parallel to $\bk$
in the weak potential limit is well explained and reproduced by
second order perturbation theory (Supplementary Discussion~2).

Remarkably, the anisotropy in energy dispersions
of the 1D superlattices can be tuned by changing
the applied potential in such a way that $v_\perp$
is reduced completely to zero (Fig.~2c).
Hence, we can achieve extremely low mobility in one direction
and normal conduction in another one simultaneously.
This enables one to control the flow of electrons dramatically.
It also provides a novel non-destructive pathway to make graphene
nanoribbons~\cite{son:2006GNR_LDA,son:2006GNR_Halfmetal,han:2007PRL_GNR_bandgap,chen:2007condmat_GNR_bandgap}
which have been actively pursued by way of cutting graphene
sheets~\cite{han:2007PRL_GNR_bandgap,chen:2007condmat_GNR_bandgap}.
The chiral nature of the states in
graphene also plays a decisive role here.
In the model without chirality as discussed before
the (isotropic) group velocity of charge carriers
is reduced monotonically and never reaches zero
within a conceivable range of the potential amplitude (Fig. 2c).
We can also achieve vanishing group velocity in one direction by
changing the length parameters of the superlattice (Fig. 2d).

  \begin{figure}
  \includegraphics[width=1.0\columnwidth]
  {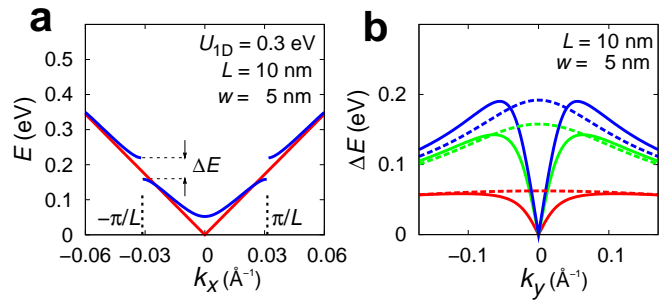}
  \caption{{\bf Energy gap at the superlattice Brillouin zone or minizone boundary of a 1D graphene superlattice.}
  {\bf a}, Energy of charge carriers in 1D graphene superlattice
  versus the component of the wavevector $\bk$ parallel to the periodic
  potential direction ($k_x$) at a fixed $k_y$.
  Dashed vertical lines indicate minizone boundaries ($k_x=\pm\pi/L$).
  $\Delta E$ is the energy gap at the minizone boundary for a given $k_y$.
  Red and blue lines correspond to $k_y$
  being zero and 0.012~\AA$^{-1}$, respectively.
  {\bf b}, $\Delta E$ versus
  $k_y$ for charge carriers in 1D graphene superlattice (solid lines)
  and that in a superlattice made from a fictitious
  system with states without chirality
  but otherwise identical to graphene (dashed lines).
  Red, green and blue lines correspond to $U_{\rm 1D}$ being
  0.1~eV, 0.3~eV and 0.5~eV, respectively.}
  \label{Fig3}
  \end{figure}

Graphene superlattices show peculiar behavior of gap openings
at the MB formed by the external periodic potential (Fig.~3).
In conventional layer-structured 1D superlattices,
gap opening at the MBs is considered to be nearly constant,
independent of $\bk$.
1D graphene superlattices, however, are different in that the gap ($\Delta E$)
vanishes when $\bk$ is along the direction of the periodic potential, i.e.,
at the centre of the MB (Fig.~3a and 4a). Moreover,
the size of the gap depends strongly on where it is on the MB (Fig.~3b).
These strong anisotropies in the gap opening do not happen
in superlattices made from a system
having linear energy dispersions but no chirality (Fig.~3b).
Hence, again, the chiral nature of charge carriers in graphene 
is key in generating these anisotropies in the gap opening
as it does in the velocity renormalization.
In particular, the gap closure at the centre of the MB
is directly related to the absence of back-scattering
of charge carriers from a scattering potential
when the size of the scatterer is several times larger than
the inter-carbon distance~\cite{ando:1998JPSJ_NT_Backscattering,
ando:1998JPSJ_NT_BerryPhase,mceuen:1999PRL_NT_Backscattering}.
In 1D graphene superlattices,
the important length-scale is $L$,
which is much larger than
the inter-carbon distance, and hence the gap does not open
at the centre of the MB.

The largest gap at the MB in a graphene superlattice is
proportional to the amplitude of the applied potential
if the potential is weak (i.e., small compared to the band width)
and the size of which thus can be made to be
a few tenths of an electron volt
with appropriate perturbation and much larger than
room temperature (Fig.~3b, Supplementary Discussion~5).
We have also investigated the gap opening in 1D graphene superlattices
with different values of length parameters ($L$ and $w$).
We find that, by changing these parameters,
the anisotropy in the gap at the MB can also be controlled
(Supplementary Discussion~5).

  \begin{figure}
  \includegraphics[width=1.0\columnwidth]
  {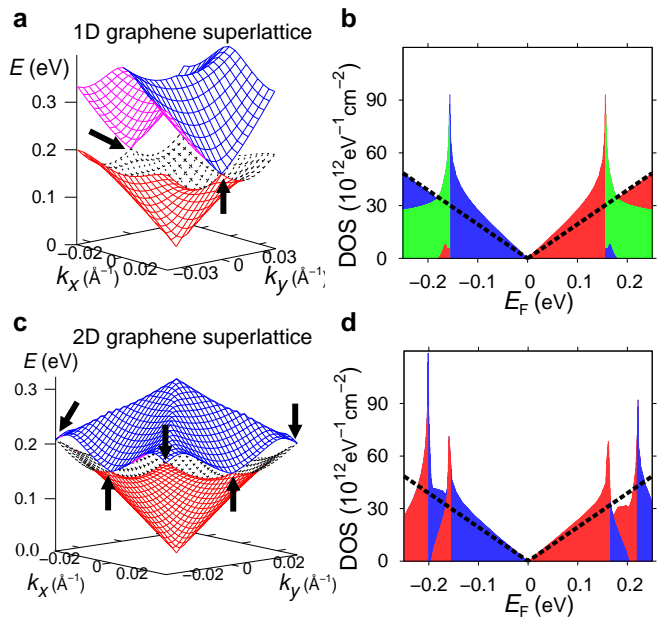}
  \caption{{\bf Energy dispersions and densities of states of charge carriers in graphene superlattices.}
  {\bf a}, Energy of charge carriers in 1D graphene superlattice
  with $U_{\rm 1D}=0.3$~eV, $L=10$~nm and $w=5$~nm in the first
  (red and black) and the second (blue and pink) band above
  the vertex of the Dirac cone
  versus 2D wavevector $\bk$ away from the Dirac point.
  Minizone boundaries are at $k_x=\pm0.031$~\AA$^{-1}$.
  Arrows indicate points on the minizone boundary where the gap closes.
  {\bf b}, Density of states (DOS) of charge carriers
  in electron orbits (red), open orbits (green) and hole orbits (blue)
  in the 1D graphene superlattice characterized in {\bf a}
  versus the Fermi energy ($E_{\rm F}$).
  The origin of the energy scale is set at the energy of the Dirac point.
  The DOS of each species is the height of
  the corresponding colored region.
  Dashed black line shows the DOS of pristine graphene for comparison.
  {\bf c}, Similar quantities as in {\bf a}
  for a 2D graphene superlattice with $U_{\rm 2D}=0.3$~eV,
  $L_x=L_y=10$~nm and $d=5$~nm.
  {\bf d}, Similar quantities as in {\bf b}
  for the 2D graphene superlattice specified in {\bf c}.}
  \label{Fig4}
  \end{figure}

Due to the velocity renormalization near the Dirac point
and the strong anisotropy in energy
dispersion close to the MB,
the type and the density of states (DOS) of charge carriers
vary drastically from those in graphene
as one varies the Fermi energy (Fig.~4).
As the Fermi level changes, the topology of the Fermi surface
also exhibits a dramatic variation (Supplementary Discussion~6).
For example, as the Fermi level increases from
the energy at the Dirac point,
the charge carriers of a 1D graphene superlattice
fill electron orbits and show a linear increase
in the DOS with slope larger than that of graphene;
but above certain value, the DOS of electron orbits vanishes
and charge carriers suddenly fill open orbits and hole pockets.
When the Fermi level increases further, charge carriers
are in purely open orbits and then the DOS of
electron orbits starts to reappear and increases again (Fig.~4b).
We expect that the Fermi level in a graphene superlattice
can be tuned as in graphene by applying a gate
voltage~\cite{novoselov:2005Nat_Graphene_QHE,zhang:2005Nat_Graphene_QHE,
huard:2007PRL_Graphene_Transport,ozyilmaz:2007PRL_Graphene_Transport}.
Hence, by exploiting the various characteristics of charge carriers
and the Fermi surface topology, one can manipulate a variety of
physical properties
dominated by the Fermi surface, such as conductance or
magnetoresistance, significantly. 

The anisotropic gap opening at the MB and the dramatic variation of the
characters of charge carriers with the Fermi energy
are also common in 2D graphene superlattices.
The gap at the centres of the zone boundaries closes
as in 1D graphene superlattice (Fig.~4c).
However, the gap at the corners of the 2D MB also disappears.
This behavior,
which occurs in rectangular 2D graphene superlattices in general,
has again its origin in the chiral nature
of charge carriers in graphene (Supplementary Discussion~4).
In a square 2D graphene superlattice, charge carriers are
electrons, holes or a mixture of the two depending on the Fermi level (Fig.~4d).
For general rectangular 2D graphene superlattices, charge carriers
can also be in open orbits.

Here we have presented several novel
physical properties of graphene superlattices with
Kronig-Penney type 1D and muffin-tin type 2D potentials.
Through additional calculations, we have
confirmed that all the salient
features of our findings are the same
in sinusoidal or Gaussian
types of graphene superlattices in general as well.
The novel properties discovered in the present
study thus should be of interest to the fundamental study
and practical applications of graphene systems in general.

Finally, since the massless Dirac fermions in graphene superlattices 
have some features in common with high-energy relativistic particles
propagating in anisotropic space
like the anisotropy in the group velocity~\cite{edwards:1963AJP_deformedSR},
interesting physics of the latter may also
be investigated by table-top experiments based
on our theoretical findings.

\vspace{1cm}
{\bf Acknowledgements}
C.-H.P. thanks J. D. Sau for fruitful discussions.
This research was supported by the National Science
Foundation (NSF) and by the Director, Office of Science, Office of Basic Energy
Science, Division of Material Sciences and Engineering, US Department of Energy
(DOE). Computational resources have been provided by the NSF at the National
Partnership for Advanced Computational Infrastructure and by the DOE at the
National Energy Research Scientific Computing Center. Y.-W.S. was supported by
the Korea Science and Engineering Foundation grant funded by
the Korea government (MOST).

{\bf Author Information}
  The authors declare that they have no competing financial interests.
  Correspondence and requests for materials
  should be addressed to S.G.L. (e-mail: sglouie@berkeley.edu).

\pagebreak

\section*{Supplementary Discussion 1 : Effective-Hamiltonian formalism\\}
There are two carbon atoms per unit cell in graphene,
forming two different sublattices, and hence the
eigenstate of charge carriers in graphene can
be represented by a two component basis  vector.
The Brillouin zone of graphene
is a hexagon which has two inequivalent vertices,
so-called the Dirac points,
$\bK$ and $\bK'$, that cannot be connected by
reciprocal lattice vectors.
In this work, we are considering
eigenstates near $\bK$ only
as discussed in the paper.
The effective Hamiltonian for low-energy
quasiparticles of graphene
in this basis is given by
\begin{equation}
H_0(\bk) = \hbar v_0\left(
\begin{array}{cc}
0 & -ik_x-k_y\\
ik_x-k_y & 0
\end{array}
\right)\ ,
\end{equation}
where $v_0$ is the Fermi velocity and $\bk$ the small
wavevector of the quasiparticle
from the $\bK$ point in the Brillouin zone of graphene.
The energy spectrum of this Hamiltonian is $E=s\hbar v_0k$
where $s$ is $+1$ or $-1$ for an eigenstate
above or below the Dirac point energy
which is defined to be the energy zero, respectively.
Eigenstates of this Hamiltonian is given by
\begin{equation}
\left<\br|s,\bk\right>=\frac{1}{\sqrt{2}}
e^{i(\bK+\bk)\cdot\br}\left(
\begin{array}{c}
1\\
ise^{i\theta_{\bk}}
\end{array}
\right)\ ,
\end{equation}
where $\theta_\bk$ is the angle of vector $\bk$ with respect to
the $\hat k_x$ direction.
Now, when an additional periodic potential $U(\br)$ is applied to graphene,
the scattering amplitude between states are given by
\begin{equation}
\left<s,\bk\left| U(\br)\right|s',\bk'\right>
=\sum_{\bG}\frac{1}{2}\left(1+ss'e^{-i\theta_{\bk,\bk-\bG}}\right)U(\bG)\ 
\delta_{\bk',\bk-\bG}\ ,
\label{eq:scatter}
\end{equation}
where $\bG$ and $U(\bG)$
are the reciprocal lattice vector and the corresponding
Fourier component of the external periodic potential, respectively,
and $\theta_{\bk,\bk-\bG}$ the angle from $\bk-\bG$ to $\bk$.
The energy dispersions and eigenstates of the
quasiparticles in a graphene superlattice are obtained non-perturbatively
within the single-particle picture
by solving the following set of linear equations:
\begin{widetext}
\begin{equation}
(E-\varepsilon_{s,k})\ c(s,\bk) = \sum_{s',\bG}
\frac{1}{2}\left(1+ss'e^{-i\theta_{\bk,\bk-\bG}}\right)U(\bG)\ c(s',\bk-\bG)\ ,
\label{eq:linear}
\end{equation}
\end{widetext}
where $E$ is the superlattice energy eigenvalue
and $\varepsilon_{s,\bk}=s\hbar v_0k$
the energy of the quasiparticles before applying the periodic potential.
$c(s,\bk)$ and $c(s',\bk-\bG)$ are the amplitudes of
mixing among different unperturbed quasiparticle states.

\section*{Supplementary Discussion 2 : Velocity renormalization near the Dirac point from second order perturbation theory\\}
When the external potential is weak, perturbative calculations
can give results
in good agreement with the full calculation
and also in physically more intuitive forms.
For pristine graphene, the group velocity of states
near the Dirac point is parallel to $\bk$ and of constant
magnitude ($v_0$).
For a graphene superlattice,
the renormalization in the component
of the group velocity of quasiparticles
parallel to the wavevector $\bk$ [$v_{\hat k}\equiv\bv(\bk)\cdot\hat k$]
around the Dirac point
can be obtained within second order perturbation approximation as
\begin{equation}
\frac{v_{\hat k}-v_0}{v_0}=-\sum_{\bG\neq{\bf0}}
\frac{2|U(\bG)|^2}{v_0^2|\bG|^2}{\rm sin}^2\ \theta_{\bk,\bG}\ ,
\label{eq:vel}
\end{equation}
where $\theta_{\bk,\bG}$ is the angle from $\bG$ to $\bk$.
From Eq.~(\ref{eq:vel}), it is clear that for
weak potentials, the velocity renormalization
grows as square of the amplitude of the external potential
and $v_{\hat{k}}$ depends only on the direction of $\bk$.
Throughout this document, by weak potential we mean that the condition,
\begin{equation}
\frac{|U(\bG)|}{v_0|\bG|}\ll1\ ,
\label{eq:weak}
\end{equation}
is satisfied for all the $U(\bG)$ components.
In the absence of chirality of the states, the scattering matrix element
[Eq.~(\ref{eq:scatter})] should be changed into
\begin{equation}
\left<s,\bk\left| U(\br)\right|s',\bk'\right>
=\sum_{\bG}U(\bG)\ 
\delta_{\bk',\bk-\bG}\ .
\label{eq:scatter_nonchiral}
\end{equation}
Using Eq.~(\ref{eq:scatter_nonchiral}), the similar quantity
as in Eq.~(\ref{eq:vel})
for nonchiral massless Dirac fermions is
now given by
\begin{equation}
\left(\frac{v_{\hat k}-v_0}{v_0}\right)_{\rm non-chiral}=-\sum_{\bG\neq{\bf0}}
\frac{2|U(\bG)|^2}{v_0^2|\bG|^2}\ ,
\label{eq:vel_nonchiral}
\end{equation}
which is isotropic independent of the direction of $\bk$.
Comparing Eq.~(\ref{eq:vel}) and Eq.~(\ref{eq:vel_nonchiral}),
the renormalization of group velocity in a one-dimensional (1D)
superlattice for
a fictitious graphene with states without chirality
corresponds to the maximum renormalization in the corresponding
1D graphene superlattice. This trend agrees with the results from
the full calculation when the potential is weak (Fig.~2 of paper).

If Eq.~(\ref{eq:vel}) is applied to the Kronig-Penney type of 1D
graphene superlattice periodic along the $\hat x$ direction
as discussed in the paper (Fig.~1b of paper),
\begin{equation}
\frac{v_{\hat k}-v_0}{v_0}=
-\left\{\frac{U_{\rm 1D}^2L^2}{\pi^4v_0^2}
\sum_{n>0}\frac{1}{n^4}\ {\rm sin}^2\ \left(\frac{\pi w}{L}n\right)\right\}\
 {\rm sin}^2\ \theta_{\bk,\hat x}\ ,
\label{eq:vel1D}
\end{equation}
where $L$ is the spatial period of the potential, and
$U_{\rm 1D}$ and $w$ are the height and the width of the
rectangular potential barrier,
respectively. Equation~(\ref{eq:vel1D}) is in good agreement with
the full calculation when the potential is weak (Supplementary Fig.~1)
and also shows sinusoidal behavior with respect to the polar angle
of $\bk$ with respect to the periodic direction of the potential
as well as the absence of renormalization for a particle
moving across the potential barriers.

  \begin{figure}
  \includegraphics[width=0.6\columnwidth]
  {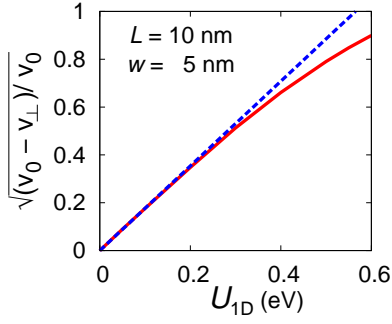}
  \caption{{\bf (Supplementary Figure 1) Dependence of the velocity renormalization on the amplitude of periodic potential in a 1D graphene superlattice.}
  Square root of the difference between the group velocity
  for state with $\bk$
  along the direction perpendicular to the periodic direction
  of the potential ($v_{\perp}$) and the
  unrenormalized one ($v_0$) divided by $v_0$ versus the potential amplitude $U_{\rm 1D}$.
  Solid red line and dashed blue line are results of the full calculation and
  second order perturbation theory, respectively.}
  \label{Fig5}
  \end{figure}

For the case of muffin-tin type of two-dimensional (2D)
graphene superlattice (Fig.~1c of paper)
periodic along both $\hat x$ and $\hat y$ directions,
with periods $L_x$ and $L_y$, respectively,
\begin{equation}
\frac{v_{\hat k}-v_0}{v_0}=
-\frac{2\pi^2U_{\rm 2D}^2\ d^2}{v_0^2L_x^2L_y^2}
\sum_{\bG\neq{\bf0}}\frac{1}{G^4}\ J_1^2\left(\frac{Gd}{2}\right)
 {\rm sin}^2\ \theta_{\bk,\bG}\ ,
\label{eq:vel2D}
\end{equation}
where $U_{\rm 2D}$ is the height of the circular potential barrier
with diameter $d$ and
$\bG=\left(\frac{2\pi}{L_x}m,\frac{2\pi}{L_y}n\right)$
the reciprocal lattice vector with integers $m$ and $n$
and $J_1(x)$ is the first order Bessel function of the first kind.
When $L_x\approx L_y$ (as in Fig.~2b of paper),
the dominant contribution in the sum comes from the terms with
$|m|=1,n=0$ and $m=0,|n|=1$. Counting only these four terms,
\begin{widetext}
\begin{eqnarray}
\frac{v_{\hat k}-v_0}{v_0}&\approx&
-\frac{U_{\rm 2D}^2\ d^2}{4\pi^2v_0^2}
\left\{\left(\frac{L_x}{L_y}\right)^2\ J_1^2\left(\frac{\pi d}{L_x}\right){\rm sin}^2\ \theta_{\bk,\hat x}+
\left(\frac{L_y}{L_x}\right)^2\ J_1^2\left(\frac{\pi d}{L_y}\right){\rm sin}^2\ \theta_{\bk,\hat y}\right\}
\nonumber\\
&=&-\frac{U_{\rm 2D}^2\ d^2}{4\pi^2v_0^2}
\left\{\left[\left(\frac{L_x}{L_y}\right)^2\ J_1^2\left(\frac{\pi d}{L_x}\right)
-\left(\frac{L_y}{L_x}\right)^2\ J_1^2\left(\frac{\pi d}{L_y}\right)\right]\right.
{\rm sin}^2\ \theta_{\bk,\hat x}\nonumber\\
&&+\left.\left(\frac{L_y}{L_x}\right)^2\ J_1^2\left(\frac{\pi d}{L_y}\right)\right\}\ 
\label{eq:vel2D_4terms}
\end{eqnarray}
\end{widetext}
where the relation ${\rm sin}^2\ \theta_{\bk,\hat y}=1-{\rm sin}^2\ \theta_{\bk,\hat x}$
has been used in the second line.
Now the group velocity $v_{\hat k}$ is renormalized in every direction.
Equation~(\ref{eq:vel2D_4terms}) reproduces the sinusoidal variation of the velocity
renormalization as well as the constant shift as obtained in the full calculation
quite well when the potential is weak (Fig.~2b of paper).

\section*{Supplementary Discussion 3 : The magnitude and the component parallel to the wavevector $\bk$ of the group velocity\\}
The component of the group velocity $v_{\hat k}$ parallel to the wavevector $\bk$
above is exactly equal to the absolute value of
the group velocity $v_g$ when
$\bk$ is at 0, 90, 180 or 270 degrees from
the periodic direction of the applied potential
and, when the applied potential is weak,
is only slightly different from $v_g$
at other angles (Supplementary Fig.~2).

  \begin{figure}
  \includegraphics[width=0.6\columnwidth]
  {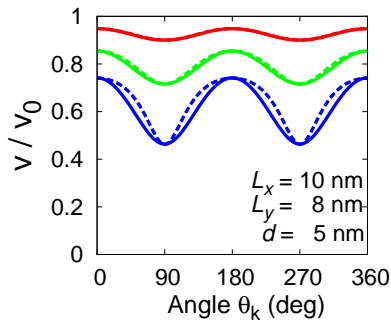}
  \caption{{\bf (Supplementary Figure 2) The magnitude and the component parallel to the wavevector $\bk$ of the renormalized group velocity in a 2D graphene superlattice.}
  The component of the group velocity
  parallel to the $\bk$ vector [$v_{\hat k}\equiv\bv(\bk)\cdot\hat k$
  with $\bk$ measured from the Dirac point]
  of charge carriers in a 2D graphene
  superlattice
  (solid lines) and the absolute value of the group
  velocity (dashed lines)
  in units of the Fermi velocity in graphene ($v_0$)
  versus the angle ($\theta_\bk$) of the $\bk$-vector
  from the periodic potential direction $\hat x$.
  Red, green and blue lines correspond to the potential
  amplitude $U_{\rm 2D}$ being
  0.3~eV, 0.5~eV and 0.7~eV, respectively. Plotted quantities
  are obtained from the full calculation by solving Eq.~(\ref{eq:linear}).}
  \label{SFig2}
  \end{figure}

\section*{Supplementary Discussion 4 : Band gap at the minizone boundary from degenerate perturbation theory\\}

When the wavevector $\bk$ is on the minizone boundary (MB) of the 1D graphene
superlattice, $k_x=\pm\pi/L$, two states $\left|s,\bk\right>$ and
$\left|s,\bk-(2\pi/L,0)\right>$ are degenerate before applying the
periodic potential. The largest
contribution to the energy eigenvalues at the MB
comes from these two degenerate states.
Scattering amplitude between these two states is
given by Eq.~(\ref{eq:scatter}).
By choosing the origin at the centre of a potential barrier,
we can make all the Fourier components of the periodic potential
real without losing generality.
Now the Hamiltonian for the two-state system is
\begin{widetext}
\begin{equation}
H(\bk) = \left(
\begin{array}{cc}
\varepsilon_{s,\bk} & \frac{1}{2}\left(1+e^{-i\theta_{\bk,\bk-(2\pi/L,0)}}\right)\
U\left({2\pi}/{L}\right)\\
\frac{1}{2}\left(1+e^{i\theta_{\bk,\bk-(2\pi/L,0)}}\right)\ U\left({2\pi}/{L}\right)
& \varepsilon_{s,\bk-(2\pi/L,0)}
\end{array}
\right)\ ,
\label{eq:gapeq1D}
\end{equation}
\end{widetext}
where $\varepsilon_{s,\bk} = \varepsilon_{s,\bk-(2\pi/L,0)}=s\hbar v_0k$ is the
energy of the charge carrier before the external potential is applied
and $U\left(2\pi/{L}\right)$ the Fourier component of the periodic potential
whose wavevector connects the two MBs at $k_x=\pm\pi/L$.
The energy separate of the eigenvalues of Eq.~(\ref{eq:gapeq1D})
(i.e., the energy gap) is given by
\begin{equation}
\Delta E = 2\ \left|U(2\pi/L)\ {\rm sin}\ \theta_{\bk,\hat x}\right|\ ,
\label{eq:gap1D}
\end{equation}
where $\theta_{\bk,\hat x}$ is the polar angle between $\bk$ and $\hat x$.
Equation~(\ref{eq:gap1D}) clearly shows that the gap opening depends on $\bk$
and, in particular, is zero at the centre of the MB (Fig.~2 of paper),
and that, as discussed in the paper, the maximum gap opening is proportional
to the amplitude of the external potential in the weak potential regime.

If the chirality of the states in graphene is manually removed
by using Eq.~(\ref{eq:scatter_nonchiral})
for the scattering matrix element, the energy gap becomes
\begin{equation}
\Delta E_{\rm non-chiral} = 2\ \left|U(2\pi/L)\right|\ ,
\label{eq:gap1D_nonchiral}
\end{equation}
in which case the gap neither closes at the centre of the MB
nor depends sensitively on the position
along the MB in the weak potential limit (Fig.~2 of paper).

For a 2D rectangular graphene superlattice,
in which the primitive translational
lattice vectors are orthogonal,
for the same reason as in the 1D case,
the gap closes at the centres of MBs, i.e., when $\bk$ is at
$(\pm\pi/L_x,0)$ or $(0,\pm\pi/L_y$).
The more interesting case is the corner of the minizone, where
the four degenerate states
$\left|s,(\pi/L_x,\pi/L_y)\right>$, $\left|s,(-\pi/L_x,\pi/L_y)\right>$,
$\left|s,(\pi/L_x,-\pi/L_y)\right>$ and  $\left|s,(-\pi/L_x,-\pi/L_y)\right>$
mix strongly among themselves by the applied periodic potential.
If we set up a similar matrix for this case like Eq.~(\ref{eq:gapeq1D})
for the 1D graphene superlattice,
the energy eigenvalues of the matrix are
\begin{equation}
E=\varepsilon_{s,\bk}\pm\sqrt{\left|
\left[U(2\pi/L_x,0)\ {\rm sin}\ \theta_{\bk,\hat x}\right]^2
-\left[U(0,2\pi/L_y){\rm cos}\ \theta_{\bk,\hat x}\right]^2\right|}\ ,
\label{eq:eig2D}
\end{equation}
where $\bk=(\pi/L_x,\pi/L_y)$ is at the minizone corner and
each eigenvalue is doubly degenerate.
The energy spectrum given by Eq.~(\ref{eq:eig2D}) clearly shows that
there is no gap at the minizone corner between the first
and the second band (Fig.~4c of paper).
This gap closure at the minizone corner
is not obvious because a transition, which is
not of a backscattering process,
from one of the four $\bk$ points at the
zone corners to another can occur in the 
2D rectangular graphene superlattice.
For example, the state $\left|s,(\pi/L_x,\pi/L_y)\right>$
can mix with another state $\left|s,(-\pi/L_x,\pi/L_y)\right>$
by the reciprocal lattice vector $\bG=(2\pi/L_x,0)$
but the two $\bk$ vectors are not anti-parallel.
To understand the origin of the gap
closure at the minizone corner,
we repeated a similar calculation for a
2D rectangular superlattice formed of
a non-chiral system with linear energy dispersions.
In this case, the energy eigenvalues are
\begin{widetext}
\begin{eqnarray}
E_{\rm non-chiral}&=&\varepsilon_{s,\bk}+U(2\pi/L_x,2\pi/L_y)
\pm\left[U(2\pi/L_x,0)-U(0,2\pi/L_y)\right]\ ,\nonumber\\
&&\varepsilon_{s,\bk}-U(2\pi/L_x,2\pi/L_y)
\pm\left[U(2\pi/L_x,0)+U(0,2\pi/L_y)\right]\ .
\label{eq:eig2D_nonchiral}
\end{eqnarray}
\end{widetext}
According to Eq.~(\ref{eq:eig2D_nonchiral}),
there is a finite energy gap
between the first and the second band in general,
other than in accidentally symmetric cases.
Therefore, the gap closure at the minizone corner
of a 2D rectangular graphene superlattice
is a direct consequence of the chiral nature of the states
in graphene.

\section*{Supplementary Discussion 5 : Dependence of the band gap at the minizone boundary on length parameters and broken particle-hole symmetry\\}

For a Kronig-Penney type rectangular potential barrier
1D graphene superlattice,
the energy gap at the MB
can be expressed with Eq.~(\ref{eq:gap1D})
in terms of superlattice parameters as
\begin{equation}
\Delta E =\frac{2}{\pi}\ \left|U_{\rm 1D}\
{\rm sin}\left(\frac{\pi w}{L}\right)\ {\rm sin}\ \theta_{\bk,\hat x}\right|\ .
\label{eq:gap1D_KP}
\end{equation}
Thus, according to this simple degenerate perturbation theory result
if the spatial period ($L$) becomes long for a constant barrier
width ($w$) the gap should decrease, whereas if $L$ is fixed and
$w$ is increased from zero, the gap should reach a maximum
at $w=L/2$ and after that should decrease symmetrically.
The former trend is observed in the full calculation
(Supplementary Fig.~3a); however, the latter
seems not to hold in the full calculation (Supplementary Fig.~3b).
Moreover, the gap openings at the MB above
and below the Dirac cone are different (Supplementary Fig.~3),
which shows the limitation of the simple degenerate perturbation theory
because the potential is not so weak and
possesses higher Fourier components.
One thing to note is that the energy dispersion
for states above the energy of the Dirac point,
including the gap opening,
for a 1D Kronig-Penney type superlattice with width
$w=w_0$ is identical with
that  of states below the energy of the Dirac point for width
$w=L-w_0$ (Supplementary Fig.~3b, compare red and blue lines).
This symmetry can be understood following a simple argument.
If we start from a 1D Kronig-Penney superlattice with
$w=w_0$ and then change $w$ to be $w=L-w_0$ and at the same time
invert the whole potential, the resulting periodic potential
is identical to the original one other than a constant shift which
may be ignored.
Inverting the potential is effectively the same as
exchanging particles with holes.

  \begin{figure}
  \includegraphics[width=1.0\columnwidth]
  {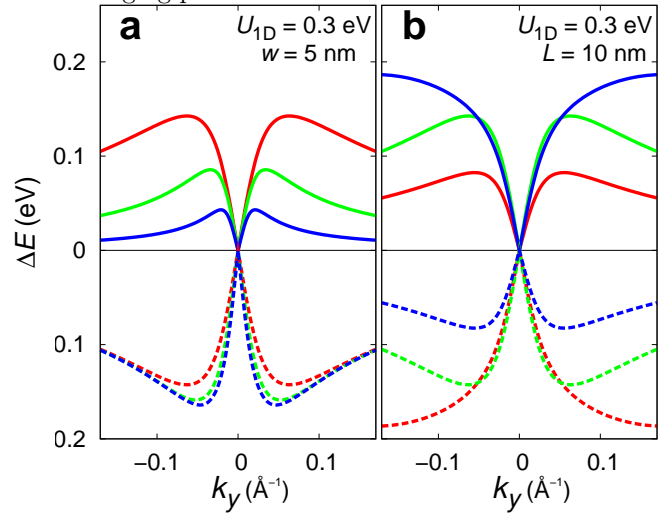}
  \caption{{\bf (Supplementary Figure 3) Dependence of the energy gap at the minizone boundary on the length parameters of 1D graphene superlattice.}
  {\bf a}, The energy gap $\Delta E$ between
  the first and the second band at the minizone boundary
  versus $k_y$ for charge carriers above (solid lines)
  and below (dashed lines) the energy at the Dirac point
  in a 1D graphene superlattice.
  Red, green and blue lines correspond to the spatial period ($L$) being
  10~nm, 15~nm and 25~nm, respectively.
  {\bf b}, Similar quantities as in {\bf a}.
  Red, green and blue lines correspond to the potential barrier
  width ($w$) being 2.5~nm, 5~nm and 7.5~nm, respectively.}
  \label{Fig6}
  \end{figure}

\section*{Supplementary Discussion 6 : Fermi surfaces\\}
In 1D and 2D graphene superlattices, the topology as well as the
shape of the Fermi surface
varies significantly with the Fermi energy (Supplementary Figs.~4 and 5).
This variation gives rise to a dramatic variation in
the species and the density of states
of charge carriers as a function of the position of
the Fermi level (Fig.~5 of paper).

  \begin{figure}
  \includegraphics[width=1.0\columnwidth]
  {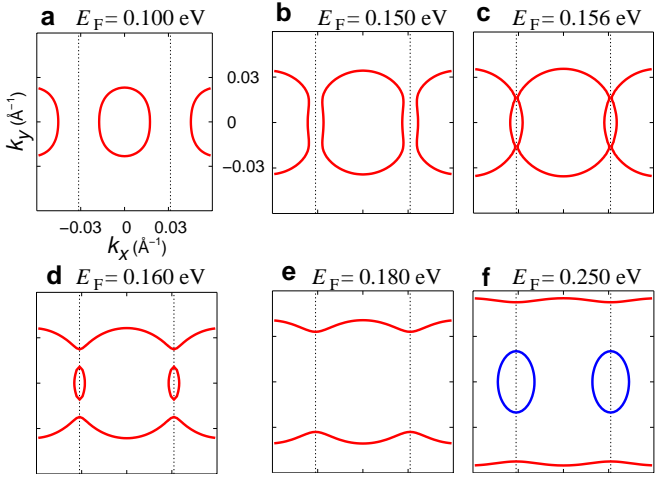}
  \caption{{\bf (Supplementary Figure 4) Fermi surfaces of a 1D graphene superlattice.}
  {\bf a-f}, Fermi surfaces of 1D graphene superlattice with
  $U_{\rm 1D}=0.3$~eV, $L=10$~nm and $w=5$~nm
  plotted in the repeated zone scheme
  for different values of the Fermi energy ($E_{\rm F}$)
  with respect to that at the Dirac point.
  Dashed lines are minizone boundaries.
  Red and blue lines are parts coming from
  the first and the second band above the Dirac point energy,
  respectively.}
  \label{Fig7}
  \end{figure}

  \begin{figure}
  \includegraphics[width=1.0\columnwidth]
  {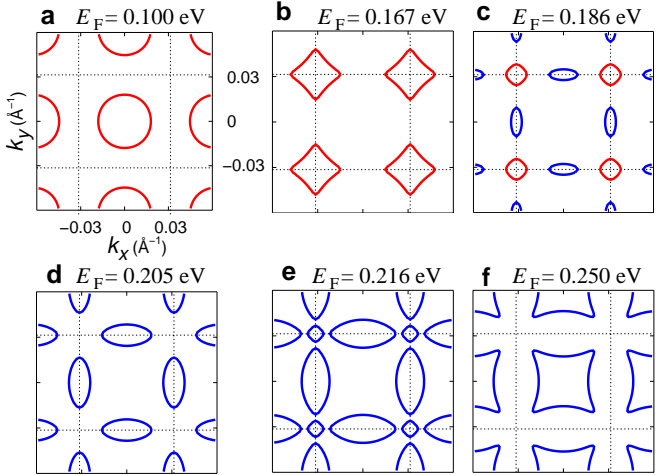}
  \caption{{\bf (Supplementary Figure 5) Fermi surfaces of a 2D rectangular graphene superlattice.}
  {\bf a-f}, Fermi surfaces of a 2D rectangular graphene superlattice with
  $U_{\rm 2D}=0.3$~eV, $L_x=L_y=10$~nm and $d=5$~nm
  plotted in the repeated zone scheme
  for different values of the Fermi energy ($E_{\rm F}$)
  with respect to the Dirac point energy.
  Dashed lines are minizone boundaries.
  Red lines and blue line are parts coming from
  the first and the second band above the Dirac point energy,
  respectively.}
  \label{Fig8}
  \end{figure}
\end{document}